\documentclass[aps,twocolumn,superscriptaddress,nofootinbib]{revtex4}

\usepackage{graphicx}
\usepackage{amssymb, amsmath,mathtools}
\usepackage{bbm}
\usepackage{enumerate}

\graphicspath{{./}{./plots/}}

\DeclareMathOperator{\Tr}{Tr}

\usepackage{xcolor}

\begin{document}

\title{Critical $O(2)$ and $O(3)$ $\phi^4$ theories near six dimensions}

\author{Igor F. Herbut}

\affiliation{Department of Physics, Simon Fraser University, Burnaby, British Columbia, Canada V5A 1S6}

\author{Lukas Janssen}

\affiliation{Department of Physics, Simon Fraser University, Burnaby, British Columbia, Canada V5A 1S6}
\affiliation{Institut f\"ur Theoretische Physik, Technische Universit\"at Dresden, 01062 Dresden, Germany}

\begin{abstract}
We consider $O(N)$-symmetric bosonic $\phi^4$ field theories above four dimensions, and propose a new reformulation in terms of an irreducible tensorial field with a cubic and Yukawa terms. The $\phi^4$  field theory so rewritten exhibits real and nontrivial IR-stable fixed points near and below six dimension, for low values of $N$ such as $N=2$ and $N=3$. The so-defined UV completions of the $O(2)$ and $O(3)$ models hence constitute precious examples of asymptotically safe quantum field theories.
The possibility of an extension of our results to five dimensions is discussed.
\end{abstract}

\maketitle

\section{Introduction}

The $\phi^4$ theory has been the cornerstone of our understanding of critical phenomena, and has served as a prototypical field theory. Its ultraviolet (UV) and infrared (IR)  properties crucially depend on the number of dimensions $d$ and the symmetry $O(N)$, where $N$ is the number of real components of the field $\phi$ \cite{wilson}. For $d<4$ the IR-stable (critical) fixed point is at a finite positive value of the self-interaction, which approaches zero as $d$ tends to four, and becomes negative and bicritical for $d>4$. At dimensions $d>4$ therefore the long-distance (IR) behavior of the $\phi^4$ theory is trivial, but there is a nontrivial, interacting, UV-stable fixed point, albeit at a negative value of the self-interaction, at which the theory appears unlikely to be completely stable. A reformulation in which the theory at the interacting  UV fixed point would emerge from the renormalization group (RG) flow of a more complete theory would therefore be useful. Indeed, such a correspondence between the UV-stable fixed point of one theory and the IR-stable fixed point of another has been established before between the Gross-Neveu model and the Gross-Neveu-Yukawa field theory for dimensions $2<d<4$ \cite{hasenfratz, zinnjustin, braun, sonoda, vacca}.

A possible such ``UV completion'' of the $\phi^4$ theory was recently proposed by Fei, Giombi, and Klebanov \cite{fei} in the form:
\begin{equation} \label{eq:lagrangian-scalar}
L= \frac{1}{2} (\partial_\mu z)^2 +\frac{1}{2} (\partial_\mu \phi_i)^2  +  g z \phi_i \phi_i + \lambda z^3,
\end{equation}
where $i=1,\dots,N$, and the summation over the repeated indices is assumed. The new scalar $z$ may be understood as the Hubbard-Stratonovich field used to decouple the original $ (\phi_i \phi_i)^2$ quartic term in the {\it scalar} channel, which has acquired its own dynamics by the integration over the high-energy modes of the original field $\phi$. The $O(N)$ symmetry also allows two IR-relevant quadratic terms, $m^2 _z z ^2$, and  $m^2 _\phi \phi_i \phi_i$, which have been  tuned to zero to be at the critical surface. The advantage of this reformulation of the $\phi^4$ theory is that the two interaction coupling constants allowed by the $O(N)$ symmetry, the self-interaction of the new fields $z$, $\lambda$, and the ``Yukawa'' coupling $g$, are both marginal in the same dimension $d=6$. Below six dimensions therefore one may hope to find a weakly-coupled IR-stable fixed point at infinitesimal values of $g$ and $\lambda$, when $m_z =m_\phi=0$. Such an interacting $O(N)$-symmetric field theory above four dimensions would then at large distances correspond to the $\phi^4$ theory at the UV-stable fixed point in the original formulation, and provide a stable, conformal, UV-complete version of it.

The above correspondence was further argued \cite{fei, nakayama, chester} to be indeed born out at a sufficiently large value of $N$ \cite{percacci}: to the lowest order in the small parameter $\epsilon = 6-d $ the fixed-point values of the couplings $g$ and $\lambda$ are real, and consequently the theory is unitary, only for $N > 1038$. Further two-loop and three-loop corrections \cite{fei2} indicated a possible dramatic reduction of the critical value of $N$ when the parameter $\epsilon$ is extended to the physical value of $\epsilon=1$, for example, but the question of the existence of the unitary $O(N)$-symmetric conformal field theory in five dimensions for reasonably low values of $N$  has remained open. In this paper we propose a different reformulation of the $\phi^4$ theory, which yields such weakly-coupled, real, $O(N)$-symmetric IR-stable fixed points for $N=2$ and $N=3$ near six dimensions. The idea is to decouple the $\phi^4$ term in the \emph{tensorial} instead of in the scalar channel. Instead of the theory \eqref{eq:lagrangian-scalar} we consider:
\begin{align} \label{eq:lagrangian-tensor}
L= \frac{1}{2} (\partial_\mu z_a)^2 +\frac{1}{2} (\partial_\mu \phi_i)^2 +  g z_a \phi_i \Lambda^a  _{ij}\phi_j + \lambda \Tr[( z_a \Lambda^a)^3].
\end{align}
Here the indices $i,j=1,\dots,N$ as before, but $a=1,\dots,M_N$ where $M_N = (N-1)(N+2)/2$ is the number of components of the irreducible (traceless) tensor of the second rank under $O(N)$ rotations. The $M_N$ matrices $\Lambda^a$ provide a basis in the space of traceless, real, symmetric  $N$-dimensional matrices, and reduce to the two familiar real Pauli matrices for $N=2$, and the five real Gell-Mann matrices for $N=3$, for example \cite{janssen}. The two IR-relevant mass terms, $m^2 _z z_a z_a$ and $m_\phi ^2 \phi_i \phi_i$, have again been tuned to zero.  Again, at such a critical surface with $m_z=m_\phi=0$ the couplings $\lambda$ and $g$ are the only ones that at the Gaussian fixed point turn IR relevant infinitesimally below $d=6$, and one hopes for a weakly-coupled IR-stable fixed point at real values of the couplings. We indeed find such IR-stable fixed points of \eqref{eq:lagrangian-tensor} for $N=2$ and $N=3$ near $d=6$ in our one-loop calculation, but not for $N \geq 4$. Our approach  may therefore be understood as being complementary to the large-$N$ strategy  of Ref.~\cite{fei}. The situation becomes particularly simple and transparent when $N=2$ and the cubic term in $\phi_a$ in fact vanishes identically, leaving the theory with a single IR-relevant Yukawa coupling $g$. We will therefore begin our discussion in the next section with this example.

The paper is organized as follows. In the next section we start with the simplest example when $N=2$. The general case is then discussed next, in Sec.~III. In Sec.~IV we present the one-loop flow equations and the concomitant fixed-point structure in the $\lambda$-$g$ critical plane. We briefly summarize and discuss the results in the concluding section.

\begin{figure*}
 \includegraphics[width=0.23\textwidth]{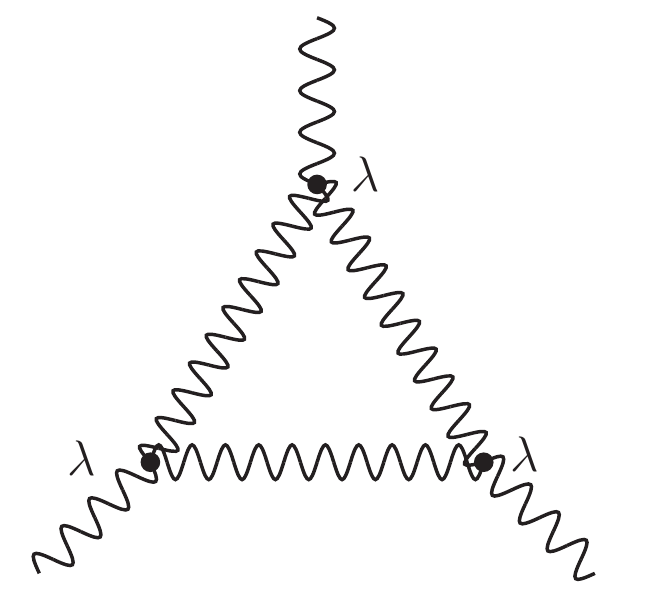}
 \includegraphics[width=0.23\textwidth]{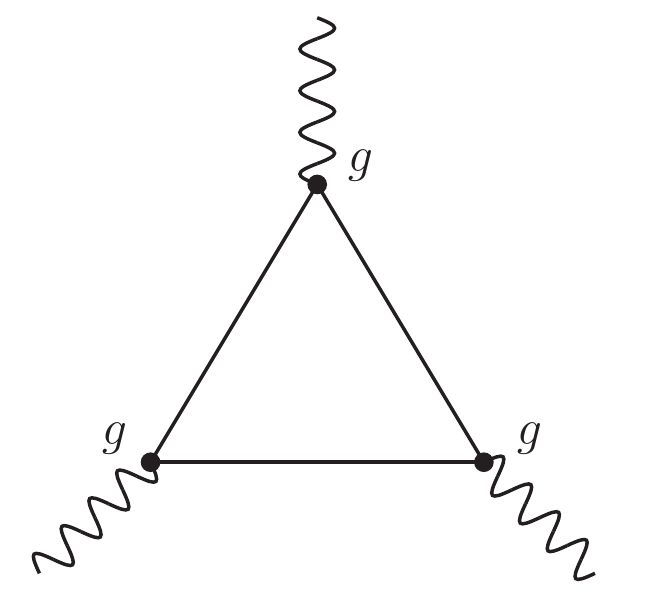}\hfill
 \includegraphics[width=0.23\textwidth]{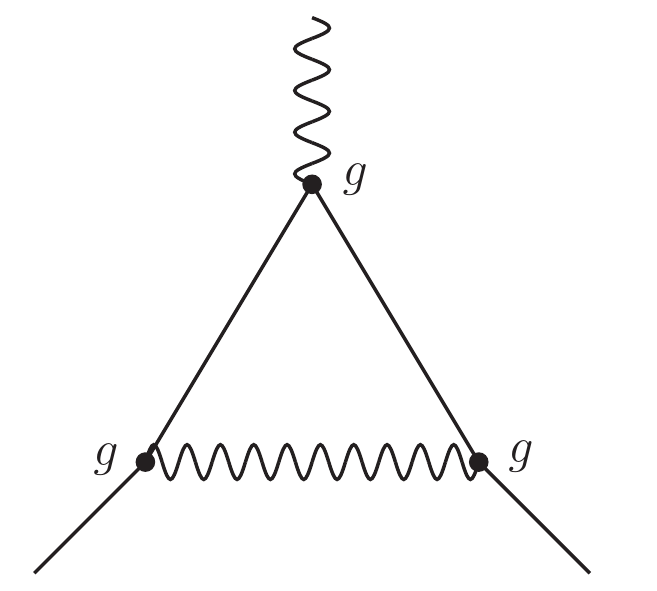}
 \includegraphics[width=0.23\textwidth]{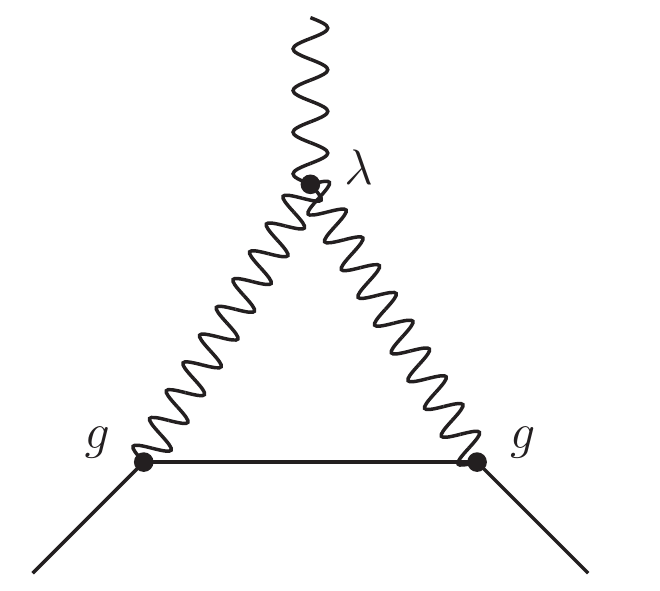}\\[\baselineskip]
 \includegraphics[width=0.225\textwidth]{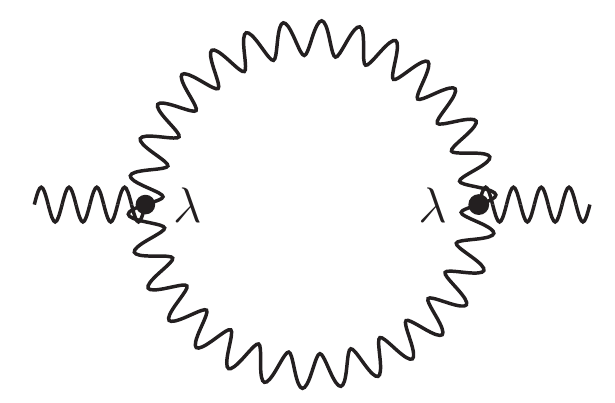}
 \includegraphics[width=0.245\textwidth]{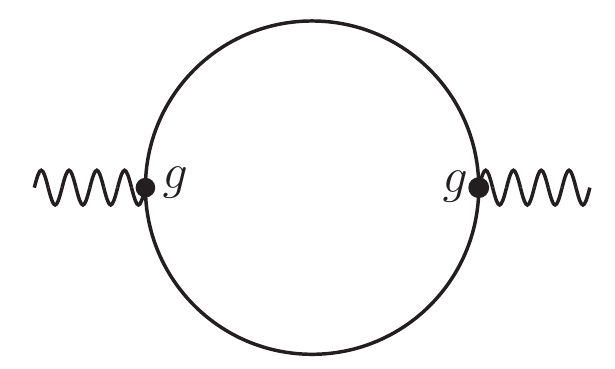}\hfill
 \raisebox{1.5\baselineskip}{\includegraphics[width=0.23\textwidth]{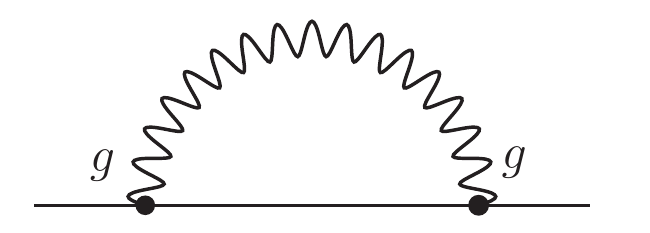}}\hfill\phantom{.}
 \caption{The one-loop diagrams that contribute to the RG flow $d\lambda/d \ln b$ (top left) and $d g/d \ln b$ (top right) and anomalous dimensions of tensor field $\eta_z$ (bottom left) and of scalar field $\eta_\phi$ (bottom right). Wiggly internal lines: tensor-field propagator~($z$). Solid internal lines: scalar-field propagator~($\phi$).}
 \label{fig:diagrams}
\end{figure*}

\section{ $O(2)$  theory}

Consider the bosonic $\phi^4$ field theory with two-component real field $\phi^T =(\phi_1, \phi_2)$. The quartic term can obviously also be written as
\begin{align}
\left(\phi_1 ^2 + \phi_2 ^2\right)^2 & =  \left(\phi_1 ^2 - \phi_2 ^2 \right)^2  + \left(2 \phi_1 \phi_2 \right)^2 \\ \nonumber
& = \left(\phi^T \sigma_{3} \phi \right) ^2 +  \left(\phi^T \sigma_{1} \phi \right) ^2.
\end{align}
This suggests the following alternative Hubbard-Stratonovich decoupling of the negative quartic term in the Lagrangian density,
\begin{equation}
L_0= \frac{1}{2} \phi_i (m_\phi ^2 - \partial_\mu ^2  ) \phi_i  - \frac{g^2}{2} \left(\phi_1 ^2 + \phi_2 ^2\right)^2,
\end{equation}
which is
\begin{equation} \label{eq:hubbard-stratonovic-2}
- \frac{g^2}{2} \left(\phi_1 ^2 + \phi_2 ^2\right)^2 = \frac{1}{2} z_a z_a + g z_a  \phi^T \sigma_{a} \phi,
\end{equation}
where the index $a\in\{1,3\}$, and $\sigma_a$ are the two real Pauli matrices.
The equivalence of the left- and right-hand sides of Eq.~\eqref{eq:hubbard-stratonovic-2} is exact (modulo normalization) at the level of the partition function,
\begin{widetext}
\begin{equation} \label{eq:partition-function}
 \mathcal Z = \int \mathcal D\phi_i
 \exp \biggl[ - \int d^d x L_0 \biggr]  =
 \int \mathcal D\phi_i \mathcal D z_a
 \exp\biggl[
   -  \int d^d x \biggl(
    \frac{1}{2}\phi_i(m_\phi^2-\partial_\mu^2)\phi_i + \frac{1}{2} z_a z_a + g z_a  \phi^T \sigma_{a} \phi
    \biggr)
 \biggr],
\end{equation}
\end{widetext}
which can straightforwardly be verified by integrating out $z_a$ on the right-hand side of Eq.~\eqref{eq:partition-function}. The integration over the high-energy modes of the field $\phi$ will generate further terms in the expansion in powers of the auxiliary field  $z_a$ and the momentum, as allowed by the $O(2)$ symmetry, such as $(\partial_\mu z_a)^2$,  $(\partial_\mu z_a)^4$, $(z_a z_a )^2$, etc.
Note that for $N=2$ no term cubic in $z$ is possible, as $\Tr(\sigma_a \sigma_b \sigma_c) = 0$ for $a,b,c \in \{1,3\}$.
Among the above allowed terms, in $6-\epsilon$ dimensions we need to keep only the first one, since all others will be IR irrelevant near the putative weakly-coupled fixed point. The field theory therefore becomes
 \begin{align}
L= \frac{1}{2} z_a (m_z ^2 - \partial_\mu ^2  ) z_a +  \frac{1}{2} \phi_i ( m_\phi ^2  -\partial_\mu ^2) \phi_i  + g z_a \phi^T \sigma_a \phi,
\end{align}
which is recognized to be a special case of Eq.~\eqref{eq:lagrangian-tensor}, in which  the term cubic in $z$ is absent, and with the two mass terms explicitly displayed for completeness.

\section{ $O( N>2 )$ theory }

To see that a similar trick as in the previous section can be played for any $N$, consider the quadratic form with the Hubbard-Stratonovich fields $z_a$, $a=1,\dots,M_N$, and the basis in the traceless, real, symmetric $N$-dimensional matrix space $\Lambda^a$. In the sense of Hubbard-Stratonivich transformation,
\begin{equation} \label{eq:hubbard-stratonovic-3}
\frac{1}{2} z_a z_a + g z_a \phi^T \Lambda^a \phi =
-\frac{g^2 }{2} \phi_i \Lambda^a _{ij} \phi_j    \phi _k \Lambda^a _{kl} \phi_l.
\end{equation}
Since, on the other hand, completeness of the set of matrices $\Lambda^a$ in the space of real, symmetric, $N$-dimensional matrices implies that~\cite{janssen}
\begin{equation}
\Lambda^a _{ij} \Lambda^a _{kl}= \delta_{ik} \delta_{jl} +   \delta_{il} \delta_{jk} - \frac{2}{N} \delta_{ij} \delta_{kl},
\end{equation}
one finds that the right-hand side of Eq.~\eqref{eq:hubbard-stratonovic-3} is actually proportional to the standard $\phi^4$ term:
\begin{equation}
\frac{1}{2} z_a z_a + g z_a \phi^T \Lambda^a \phi = g^2 \left(\frac{1}{N}-1 \right) (\phi_i \phi_i)^2.
\end{equation}
This identity, which generalizes Eq.~\eqref{eq:hubbard-stratonovic-2} to higher values of $N$, implies that the usual $O(N)$-symmetric quartic term $(\phi_i \phi_i)^2$ can also be Hubbard-Stratonovich decoupled using $M_N$ real fields $z_a$ that transform under the rotational group $O(N)$ as the components of the traceless symmetric second-rank tensor. The main novelty compared to the $N=2$ case is that the cubic invariant
\begin{equation}
\Tr \left[(z_a \Lambda^a)^3  \right]
\end{equation}
is finite when $N>2$, and has the same engineering scaling dimension as the Yukawa term $z_a \phi^T \Lambda^a \phi$. In $6-\epsilon$ dimension and when $\epsilon \ll 1$ it suffices then to consider the theory in Eq.~\eqref{eq:lagrangian-tensor} with the two displayed cubic interaction terms allowed by the $O(N)$ symmetry only.

\section{RG flow}

It is relatively straightforward to compute the one-loop beta functions for the two cubic-term couplings. We set $m_z = m_\phi =0$, and consider general $N$. (It will be possible to  set $N$ to the special value $N=2$ afterwards.) Performing the usual Wilson's momentum-shell mode elimination~\cite{herbutbook}, under the change of the UV cutoff $\Omega$ into $\Omega/b$ the couplings flow according to the equations
\begin{align}
 \frac{d\lambda}{d\ln b} & = \frac{1}{2} ( \epsilon - 3 \eta _z) \lambda
 + 36 \left(N+4 - \frac{24}{N} \right) \lambda^3 +\frac{4}{3} g^3, \label{eq:flow-lambda}\\
 \frac{dg}{d\ln b} & = \frac{1}{2} ( \epsilon - \eta _z- 2\eta_\phi) g
 + 4 \left(1 - \frac{2}{N} \right) g^3
 \nonumber \\
 & \quad + 12 \left(N+2 - \frac{8}{N} \right) g^2 \lambda,
\end{align}
where the anomalous dimensions appearing above are
\begin{gather}
 \eta_z = 12 \left(N+2 - \frac{8}{N}\right) \lambda^2 + \frac{4}{3} g^2,  \\
 \eta_\phi = \frac{4}{3}  \left(N+1 - \frac{2}{N}\right) g ^2. \label{eq:eta-psi}
\end{gather}
Note that both anomalous dimensions are manifestly positive at any real fixed point and for $N\geq 2$.
Here $\epsilon = 6-d$, and
we rescaled the couplings as $\Omega^{(3\eta_z - \epsilon)/2} [ S_d / (2\pi)^d ]^{1/2} \lambda \mapsto \lambda$ and $\Omega^{(\eta_z + 2\eta_\phi - \epsilon)/2} [ S_d / (2\pi)^d ]^{1/2} g \mapsto g$.
$S_d$ is the usual surface area of the unit sphere in $d$ dimensions. The diagrams that lead to the Eqs.~\eqref{eq:flow-lambda}--\eqref{eq:eta-psi} are depicted in Fig.~\ref{fig:diagrams}.

Small perturbations out of the critical surface $m_z = m_\phi = 0$ are relevant in the sense of the RG, and governed by the flow equations
\begin{align}
 \frac{dm_z^2}{d \ln b} & = \left (2-\eta_z + 72 \left(N+2-\frac{8}{N}\right )\lambda^2 \right) m_z^2
 \nonumber \\
 &\quad + 8 g^2 m_\phi^2,
 \\
 \frac{dm_\phi^2}{d \ln b} & = (2-\eta_\phi)m_\phi^2
 \nonumber \\
 &\quad + 4 \left(N+1-\frac{2}{N}\right) g^2 (m_\phi^2 +m_z^2),
\end{align}
where we rescaled $\Omega^{\eta_z - 2} m_{z}^2 \mapsto m_{z}^2$, $\Omega^{\eta_\phi-2} m_{\phi}^2 \mapsto m_{\phi}^2$ and shifted the masses so that the position of the critical surface remains $m_{z} = m_\phi =0$.

\begin{figure}[bt]
 \includegraphics{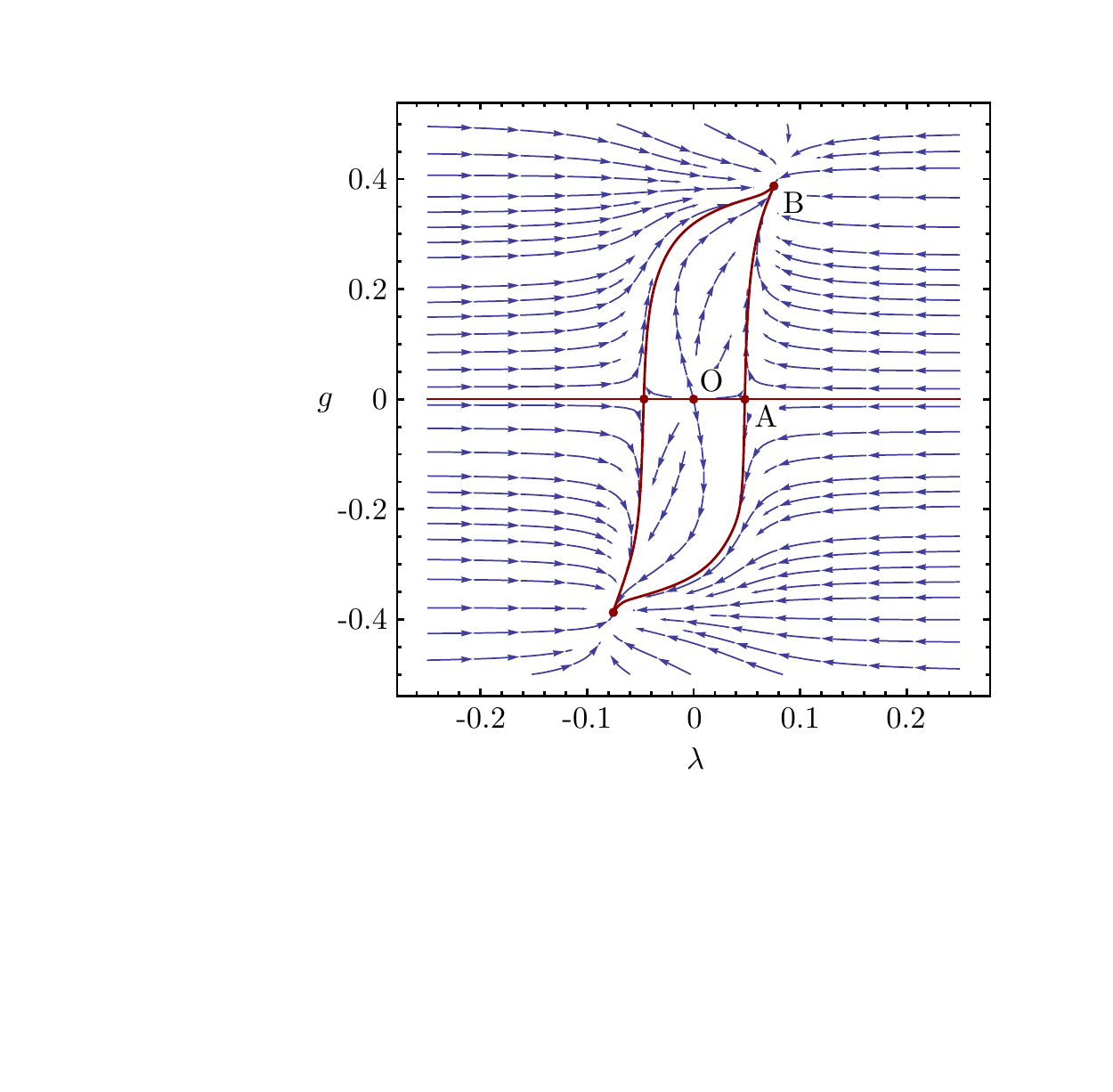}
 \caption{RG flow diagram at the critical surface $m_z = m_\phi =0$ for $1<N<2.653$. O corresponds to the noninteracting, Gaussian, fixed point. Fixed point A at $g^*=0$ and $\lambda^*>0$ is unstable towards the direction of $g$. B denotes the stable fixed point. For the physical value of $N=2$ (displayed), the term in Eq.~\eqref{eq:lagrangian-tensor} that is cubic in $z$ vanishes, and different values for $\lambda$ (horizontal axis) correspond to equivalent physical points.}
 \label{fig:flow_02}
\end{figure}

\begin{figure}[tb]
 \includegraphics{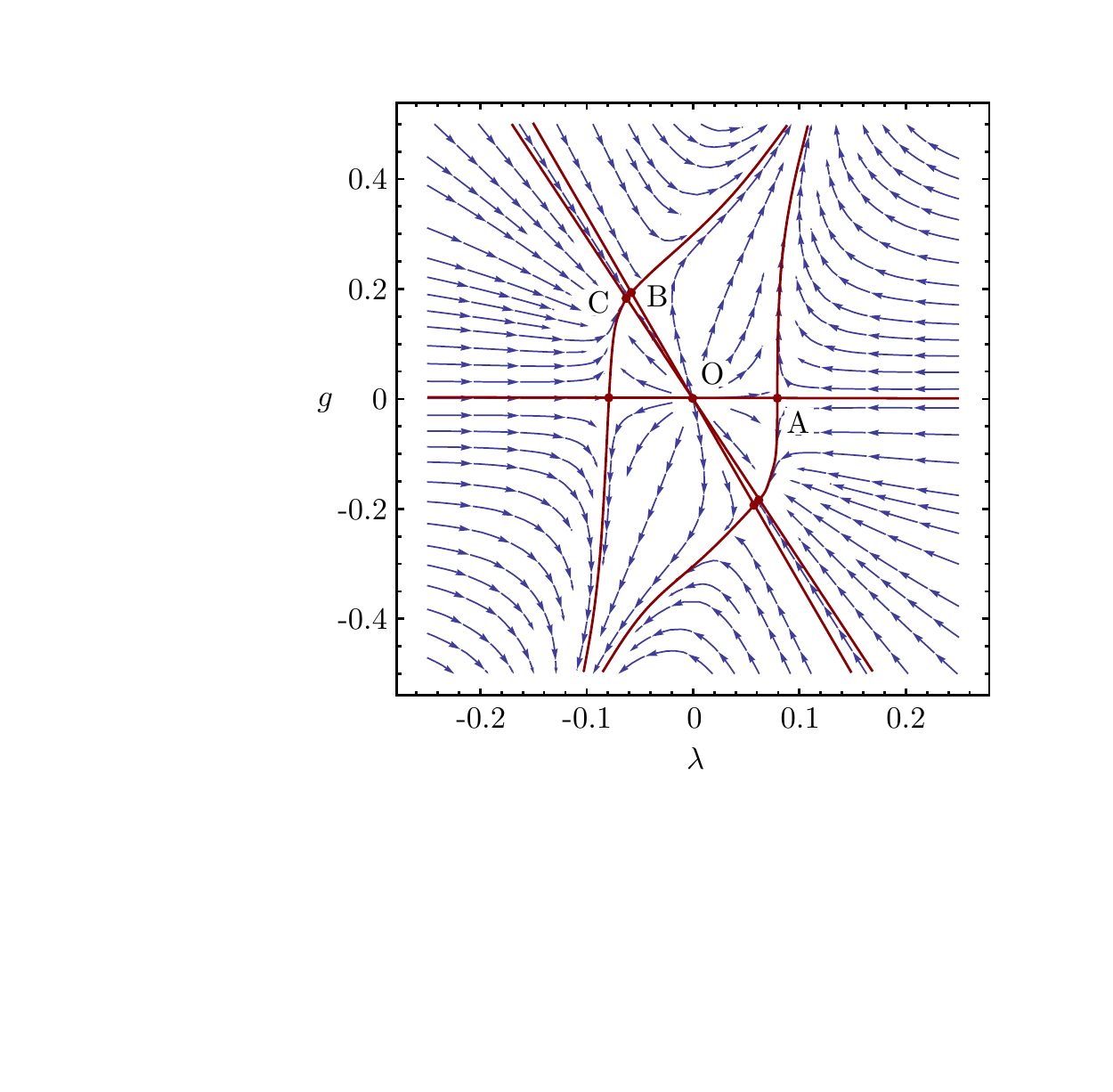}
 \caption{RG flow diagram at the critical surface, for $N=3$. The stable fixed point is denoted by C, whereas A and B are both unstable in one direction.
 The fixed points B and C are very close to each other, due to the value of $N=3$ being near the critical value of $N = 2.999$ at which they would coincide. For $N$ below this critical value the fixed points B and C would both become complex~\cite{kaplan}.}
 \label{fig:flow_03}
\end{figure}

It is interesting to consider the evolution of the fixed-point structure of these equations with $N$, treated as a continuous variable.

\begin{enumerate}[(1)]

\item
 For $1<N<2.6534$ there is stable fixed point on the critical surface $m_z = m_\phi =0$. For the physical value of $N=2$ the flow equation for $g$ as well as the anomalous dimensions become independent of $\lambda$, reflecting the fact that the term cubic in $z$ in Eq.~\eqref{eq:lagrangian-tensor} vanishes in this case. We then find the location of the fixed point along the $g$-axis to be
 \begin{equation}
 (g^*)^2 =\frac{3}{20} \epsilon.
 \end{equation}
 At this fixed point, the anomalous dimensions have the values
 \begin{equation}
 \eta_\phi = 2 \eta_z = \frac{2}{5}\epsilon.
 \end{equation}
 The RG flow in this situation is depicted in Fig.~\ref{fig:flow_02}, displaying the IR-stable fixed point B.
 The only IR-relevant parameters at this fixed point correspond to the two masses of the fields $z$ and $\phi$, and are governed by the universal exponents $\theta_{1,2}$ that are the eigenvalues of the mass-mixing matrix
 \begin{equation}
 \frac{\partial}{\partial(m_z^2, m_\phi^2)} \left(\frac{d m_z^2}{d \ln b}, \frac{d m_\phi^2}{d \ln b}\right),
 \end{equation}
 given by (for $N=2$)
 \begin{align}
 \theta_1 & = 2 + \frac{8}{5} \epsilon,
 &
 \theta_2 & = 2 - \epsilon.
 \end{align}

\item
 As $N \nearrow 2.6535$, the above fixed point B runs away to infinity. For $2.6535 < N < 2.9990$ there is no stable fixed point on the critical manifold.

\item
 For $2.9991 < N < 3.6846 $ there is a stable fixed point at the critical manifold again. At the physical value of $N=3$ it is located at
 \begin{align}
 (\lambda^*)^2 & = \frac{\epsilon}{264},
%
 &
 (g^*)^2 & = \frac{3\epsilon}{88},
 \end{align}
 and with $\lambda^*$ and $g^*$  of opposite sign. The anomalous dimensions at $N=3$ are then
 \begin{equation}
 \eta_z = \eta_\phi = \frac{5}{33}\epsilon,
 \end{equation}
 with the universal exponents corresponding to the two relevant directions out of the critical surface
 \begin{align}
 \theta_1 & = 2 + \frac{25}{33} \epsilon,
 &
 \theta_2 & = 2 + \frac{1}{33} \epsilon.
 \end{align}
 Note that the theory \eqref{eq:lagrangian-tensor} is invariant under simultaneous change of sign of $\lambda$ and $g$ \cite{janssen}. The two fixed points at $\lambda^*>0$, $g^*<0$ and $\lambda^*<0$, $g^*>0$ are therefore physically equivalent.
 Fig.~\ref{fig:flow_03} illustrates the flow for $N=3$, displaying the IR-stable fixed point~C.

\item
  For $3.6847 < N < 4$ the fixed point at $g^*=0$ and $\lambda^*>0$ (fixed point A in Figs.~\ref{fig:flow_02}--\ref{fig:flow_03}) becomes stable. This is the fixed point discussed before in the context of the thermal nematic phase transition in liquid crystals by Priest and Lubensky \cite{priest}. As $N \rightarrow 4$, however, the value of $\lambda^*$ runs off to infinity, so that for $4\leq N$ there are again no stable fixed points at the critical manifold.
\end{enumerate}

\section{Summary and open questions}

We have found that both physical values of $N=2$ and $N=3$ lie within the intervals in which there is a stable nontrivial fixed point of the theory~\eqref{eq:lagrangian-tensor}, at the critical manifold with both relevant quadratic terms tuned to zero.
 It is tempting to conjecture that these fixed points describe the same universal physics as the usual Wilson-Fisher fixed point at negative coupling in the $\phi^4$ theory above four dimensions. These fixed points allow a consistent and predictive UV completion of the theory---its perturbative nonrenormalizability notwithstanding. Near and below six dimensions our tensorial cubic theory hence constitutes another precious example of an asymptotically safe quantum field theory~\cite{weinberg1979}.
The open question, however, is whether these fixed points survive the extension all the way to five dimensions. The $N=3$ fixed point, being so close to the edge towards the region without stable fixed points (see Fig.~\ref{fig:flow_03}), seems to be in particular danger. This issue requires further study, by higher-order epsilon expansion, or functional RG, for example. One expects the critical values of $N$ at which the fixed-point structure of the flow diagram qualitatively changes to behave as \cite{zlatko}
\begin{equation}
  N_c = N_0 + N_1 \epsilon+ N_2 \epsilon^2 + O(\epsilon^3),
\end{equation}
where we have computed only the leading terms $N_0$ here to be $1$, $2.653$, $2.999$, and $4$. The corrections to these values would then follow from two-loop ($N_1$), three-loop  ($N_2$), and higher-order calculations, similarly as computed in Ref.~\cite{fei2} for the theory in Eq.~\eqref{eq:lagrangian-scalar}. The existence of the nontrivial fixed points we predict should in principle also be testable in Monte Carlo RG studies,
e.g., by employing the technique developed recently to identify the UV fixed point of the three-dimensional nonlinear sigma model~\cite{wellegehausen2014}.

While the present scheme enables us to reveal the fixed-point structure and the stability properties of the RG flow, it appears hard to deduce the full form of the effective fixed-point potential and to examine the potential's (global) stability. It should be worthwhile to reconsider this question in a future analysis, e.g., along the lines put forward in Ref.~\cite{borchardt2015}.

\begin{acknowledgments}
We are grateful to H. Gies for useful discussion. The authors acknowledge the support by the DFG under JA2306/1-1, JA2306/3-1, and SFB 1143, as well as the NSERC of Canada.
\end{acknowledgments}

\end{document}